# Nanolayered diffusion couple technique for the reconsidering of the miscibility gap

Yusuke Iguchi[1,a*], Gábor Katona[2], Csaba Cserháti[2]
Gábor Langer[2], Attila Csik[1] and Zoltán Erdélyi[2,b]

[1] Institute for Nuclear Research, Hungarian Academy of Sciences, Debrecen, Hungary
[2] Department of Solid State Physics, University of Debrecen, Debrecen, Hungary
[a] iguchi.yusuke@atomki.mta.hu  [b] zoltan.erdelyi@science.unideb.hu

**Abstract.** Mysteries in binary phase diagram are still in existence, e.g., the case in Cu-Ni which is mutual miscible system, the question addressed that "Indirect experimental evidence indicates the presence of a miscibility gap in the fcc phase at a temperature somewhere between 450 and 923 K". Recently, author's group suggests new miscibility gap on Cupper-Nickel system by means of Secondary Neutral Mass Spectroscopy (SNMS) with nanolayered diffusion couple technique. Here we considered about problems in this technique, e.g., strain and size modified miscibility gap, equilibrium and its morphorogy, how to differentiate it, etc. Finally we discussed for justification of our procedure and phase diagrams on several binary alloys.

## Introduction

Mysteries in binary phase diagram are still in existence, e.g., in the case of Cu-Ni which is mutual miscible system, the question addressed that "Indirect experimental evidence indicates the presence of a miscibility gap in the fcc phase at a temperature somewhere between 450 and 923 K." [1]. And recently several reconsidering of Cu-Ni miscibility gap using computing has reported [2,3].

Our group also suggested new miscibility gap on Cupper-Nickel system by means of Secondary Neutral Mass Spectroscopy (SNMS) with nanolayered diffusion couple technique [4].

In this study we verify the definition about the miscibility gap and discuss for justification of nanolayered diffusion couple technique with clearing for several problems, then several demonstrations are done by means of SNMS.

## Prior knowledge on the miscibility gap

**Problems on the miscibility gap.** The spinodal decomposition and its strain effect was defined finely by Cahn [5,6], ideally un-stressed one was named as "chemical spinodal" and stressed one as "coherent spinodal". Then Jantzen concluded about chemical/coherent spinodal decomposition and incoherent/coherent miscibility gap well [7] as shown in Figure 1.
Here "chemical (incoherent)" means ideal and stress free case, "coherent" means real with arbitrary stress. Cahn noted that "The coherent phase diagram is always metastable and lies within the un-stressed equilibrium phase diagram. The coherent phase diagram is a real, metastable phase diagram since it involves reversible metastable equilibrium, which is subject only to the constraint that the lattices remain continuous."

On the other hand, for example, David noted that "This corresponds to the equilibrium compositions of incoherent phases, i.e. in the absence of strain fields." [8]. These two descriptions are contradiction for the definition of miscibility gap.

These definitions, chemical and coherent, are very unclear, for example, spinodal decomposition in constrained condition which is in coherent epitaxial thin film was called as coherent spinodal by several.

Typical in metallurgy, it is generally believed that measuring of spinodal decomposition, which has diffused interface and generates nucleation depending on condition, e.g. super cooling , is only the way to determine the miscibility gap. The curve of coherent miscibility gap (hereafter "binodal (or coexistence curve)") is calculated from coherent spinodal which is experimentally obtained from the annealing of supersaturated alloy, or estimated from thermodynamic parameter. This is an indirect way to determine the coherent binodal because composition of alloying comes from inside of miscibility gap to just or around at coherent spinodal where inter diffusion coefficient is zero.

Conversely on the diffusion couple, which is no nucleation with planer interface while in isothermal treatment , is direct determination of incoherent binodal whose composition of alloying should come from outside of miscibility gap to just or around at incoherent binodal. This should be sophisticated way to determine the incoherent miscibility gap. The critical problem in spinodal decomposition from super-saturated alloy is that this shows lower spinodal temperature than ideal one during in strain induced by decomposition itself , it is real in practical material, but impossible to estimate the incoherent binodal which is same as incoherent miscibility gap. And it doesn't show clear phase separation because it should have diffused interface .

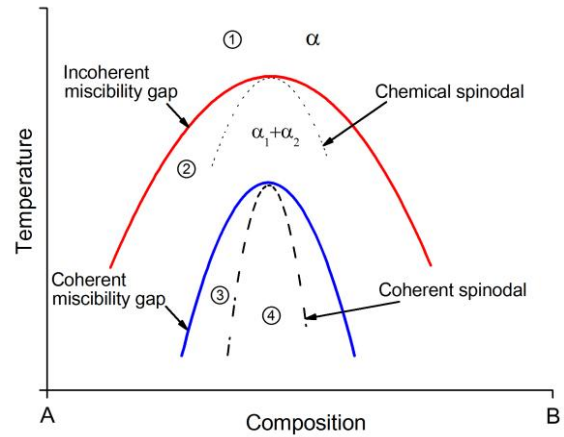

Figure. 1. Schematic phase diagram. Region 1: homogeneous a stable. Region 2: homogeneous metastable, only incoherent phases can nucleate. Region 3: homogeneous alpha metastable, coherent phase can nucleate. Region 4: homogeneous alpha unstable, no nucleation barrier, spinodal decomposition occurs, redrawn from Ref .

However we would suggest that miscibility gap should be simply solubility limits in solids with sharp interface and it have to be used as incoherent miscibility gap for the safety if it has possibilities of metastable phase.

**Diffusion couple technique for the miscibility gap.** The diffusion couple technique for the determination of phase diagram has the history of half a century. This technique was concluded well by Kodentsov and it is used for determining the phase diagram with multiple phases oftenly, e.g.. But this technique indicates different phase diagram from general ones in some complex system .

Generally in the metallurgy, there is no report about diffusion couple technique for determining the miscibility gap in pure binary alloy as far as we know, except for Kodentsov's report using pseudo binary. There is no previous experimental and detailed discussion about the binary system with miscibility gap in solids because it is impractical in bulk material at low temperature range, e.g. Cu-Ni, Ni-Pd etc. But it is demonstrated using numerical simulation by several.

Then there is previously discussion diffusion couple technique in thin film with multiple phases but no consideration for the miscibility gap as far as we know.

**Rapid diffusion for an equilibrium: A-type grain boundary diffusion and its morphology.** At first here we check again about the equilibrium on the binary system on the miscibility gap. In the case from super-saturated alloys, it is generally known that interdiffusion coefficient should be zero at spinodal, where is equilibrium . Conversely in the case from diffusion couple, diffusion stopped at binodal where is equilibrium, as explained previously .

It is difficult to be sure that is it came to the equilibrium state or not when using the "diffusion couple technique in bulk". But first and last, result in nanolayered diffusion couple technique suggests really phase separation at or very close to equilibrium cause it has enough bulk diffusion length $L_{bulk} = 2\sqrt{Dt}$ against that it is sufficiently thinner sample thickness.

As an example in the case of Cu-Ni system, $L_{bulk}$ of Nickel in Copper is about 100 nm at 722 K for 168 hours [33]. Additionally interdiffusion in this case must show type-A grain boundary diffusion because the $L_{bulk}$ is enough larger than the grain boundary width which is to use diffusion parameter from. Diffusion length, $L_{bulk}$ of Ni into Cu, as an example, is about 50 nm at 722 K for 48 hours, which is larger than the grain boundary width (0.5 nm ).

Effective diffusion coefficient of A-type grain boundary diffusion, which has an almost planar diffusion front, represents a weighted average of the volume and grain boundary diffusivity, i.e. it is much faster than bulk diffusion only. It can be simply estimated using Maxwell equation as $L_{gb}^A \approx 2\sqrt{(2/3 D_{gb})t}$ in the nano crystalline materials, which of Ni into Cu is about 120 nm at 722K for 48 hours .

Furthermore, diffusion induced recrystallization (DIR) and diffusion induced grain boundary migration (DIGM) were concerning on diffusion in the polycrystalline, but it is unnecessarily anxiousness. DIR/DIGM in thin film is found in early stage of annealing e.g. Ni/Pd , not observed in long annealing, e.g. Cu/Ni .

It should grow from grain boundary in main matrix to diffusion direction, not in A/B interface . And size of its recrystallization, e.g. , should be larger than it of nanolayered sample which has several tens nm as prepared.

Finally conclusion, enough long annealing with property temperature, previous description, also supports it that "Nanocrystalline diffusion couple must show A-type grain boundary diffusion with almost planar diffusion front".

Our result which has enough relaxing by long-time isothermal treatment does not indicate resulting DIGM/DIR. Phase separation with finally sharp interface really occurred in initial nanocrystalline thin film diffusion couple.

**Strain effect on the miscibility gaps.** As discussed earlier, any strain possibly caused a decreasing temperature of the miscibility gap, but of course there are some exceptions, e.g. strain-induced intermediate phase, etc.

Here we discuss about strain condition in our sample . We prepared bi- or tri-layered thin film diffusion couple on substrate; generally total thickness is 100 nm keeping 50-50at% or other molar ratio as a required range and polycrystalline film onto oxidized Si substrate.

Stress in general, also in our sample may originate mainly from lattice mismatch, difference in thermal expansions in bimetallic strips, and net volume flux during diffusion. Here we consider about these problems in the Cu-Ni case.

1) Lattice mismatch can be neglected because our samples are polycrystalline (non-epitaxial) and there are very small lattice mismatches during the heating CuNi compounds.

2) Bimetallic strip effect between the film and substrate can also be negligible. This is so because Young's modulus and linear thermal expansion coefficient of Cu and Ni are very close.

3) Atomic volumes of Cu and Ni are also very close, so net volume flux should not be very important; even if it is at the beginning of the intermixing, in equilibrium the stress should be relaxed.

Summarizing in stress effect, these points above the miscibility gap we measured rather corresponds to the mostly stress-free case.

**Size effect on the miscibility gap.** We should not forget about that this technique is performed in nano-scaled systems which may raise two points: the size dependence of the miscibility gap and the validity of the thermodynamic calculations on the nanoscale. The size dependence of the miscibility has been predicted in the framework of the Cahn-Hilliard concept (e.g. ). However, even Calphad experts pointed out the problem of the applicability of Calphad type thermodynamic calculations on the nanoscale . According to their nano-Calphad concept "an extension of the Calphad method for systems containing at least one phase (or at least one interface lm, complexion) with at least one of its dimensions being below 100 nm" is required; without attempting to be comprehensive: curvature dependence of the interfacial energies, dependence of interfacial energies on the separation between interfaces (including the problem of surface melting), role of the shapes and relative arrangement of phases, role of the substrate (if such exists), role of segregation.

Phase diagram and thermal stability in the nanomaterial is one of hottest researching fields recently, e.g. . They also mentioned that obvious size effect would appear in several tens nm.

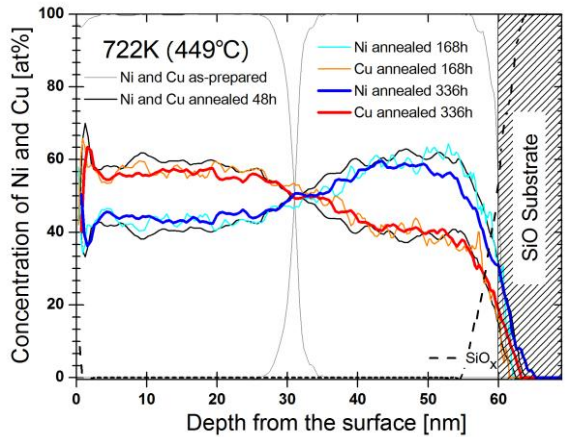 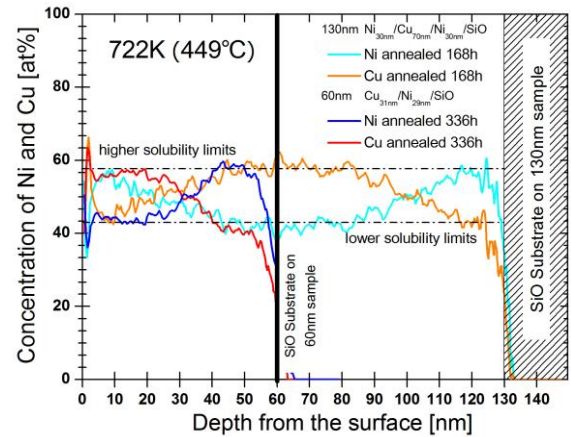

Figure. 2. Depth profiles recorded by means of SNMS of a sample $Cu_{31nm}/Ni_{29nm}/SiO_x$ (Ni composition 50 at%; total film thickness 60 nm) in as-prepared and annealed states (at 722 K for 48, 168, 336 hours). The shaded zone is the SiO_x substrate, from reference [4].

Figure. 3. Depth profiles recorded by means of SNMS of a sample $Cu_{31nm}/Ni_{29nm}/SiO_x$ annealed at 722 K for 336 hours and $Ni_{30nm}/Cu_{70nm}/Ni_{30nm}/SiO_x$ annealed at 722 K for 168 hours. Both Ni compositions are about 50at%; total film thicknesses are 60 nm and 130 nm.

**Measurement of phase separation: Secondary Neutral Mass Spectrometry**

Most of miscibility gap in binary alloy system is pair of elements with similar property, which means that it is difficult to differentiate each other by several measurements with indirect information from elements, e.g. EDX, XRD, DSC, or etc. But mass spectroscopy technique must give us direct information from elements. Therefore here we propose to demonstrate phase separating from nanolayered diffusion couple by means of Secondary Neutral Mass Spectrometry (SNMS).

SNMS using electron gas post-ionization is ideally suited to quantitative elemental and depth profiling analysis of any material. Minimal matrix effects and no influence due to preferential sputtering allow very accurate material analysis .

Here we check the bulk diffusion and A-type grain-boundary diffusion length of Cu-Ni binary alloy; Ni into Cu at 722 K for 48 hours and 168 hours. $L_{bulk}$ is about 50 nm for 48 hours and 100 nm for 168 hours, $L_{gb}^A \approx 2\sqrt{(2/3 D_{gb})t}$ is about 120 nm 48 hours and 270 nm for 168 hours. It is obviously that annealing at 722 K for 168 hours is enough annealing time to come to the equilibrium for sub-micron sample in neither case for a 60 nm sample.

Figure. 3. shows depth profiles by means of SNMS of a 60 nm sample; $Cu_{31nm}/Ni_{29nm}/SiO_x$ in as-prepared and annealed (at 722 K for 48, 168, 336 hours). The result is shown that the time coming to the mostly equilibrium state is just for 48 hours at 722 K. It is possible to declare that it should be the miscibility gap (incoherent miscibility gap) region with symmetry shape in 50-50at%. We also performed TEM experiments, but there were neither distinct phases, nor indications of DIR/DIGM.

Figure. 4. shows demonstration for evaluating the size effect between 60 nm and 130 nm, a sample of 60 nm bilayer; $Cu_{31nm}/Ni_{29nm}/SiO_x$ annealed at 722 K for 336 hours and 130 nm trilayer; $Ni_{30nm}/Cu_{70nm}/Ni_{30nm}/SiO_x$ annealed at 722 K for 168 hours. Featured note in this result is that solubility limits of both samples are mostly same in spite of a 60 nm sample is half thickness and twice annealing time than a 130 nm sample.

This resulted that there is not sufficiently size effect between 60 nm and 130 nm in Cu-Ni binary system.

Conclusions in this demonstration, several results which has plateau and stepped profile with relatively sharp interface were demonstrated by means of SNMS. From the viewpoint of theoretical study of diffusion as previous discussion , these profiles should indicate being in the miscibility gap. Also efficiently size effects are not observed around in 100 nm.

# Conclusion

Nanolayered diffusion couple technique would give us fruitful information of the miscibility gap in the solid.

■Requirement for the sample in order to come to the miscibility gap
  *Thinner thickness is better but around 100nm. Obvious size effect would appear in too thin sample (less than several tens nm).
  *Polycrystalline films are necessary for the A-type grain boundary diffusion.
  *Annealing at enough high temperature and long time for the A-type grain boundary diffusion, relaxing the strain, and ignoring the bimetallic strip when the sample with substrate.

■Cautionary note
  *Size and strain effect must decrease the critical point of the miscibility gap if it is mutual miscible system. These never raise the critical temperature.
  *Doubt the coherent miscibility gap in the case in small melting point, small Young's modulus alloy, because strain contributes possible to be decreasing temperature of the phase diagram efficiently.
  *Mass spectroscopy technique is desirable to differentiate similar elements.

   We concluded that nanolayered diffusion couple technique results in ideal solubility limits; incoherent miscibility gap. This possibly cannot apply to bulk material because it is impractical result in bulk, but this is surely ideal result with negligible strain and size effect.
This result will remind us reconsidering about the phase diagram and it of bulk and nano materials.

   The work is supported by the GINOP-2.3.2-15-2016-00041 project, which is co-financed by the European Union and the European Regional Development Fund. Y. I. acknowledges Postdoctoral Fellowship Program of the Hungarian Academy of Sciences.